\documentstyle[12pt,aasms4]{article}

\begin{document}

\title{Supermassive Objects as Gamma-Ray Bursters}

\author{George M. Fuller and Xiangdong Shi}
\affil{Department of Physics, University of California, San Diego, La
Jolla,
CA 92093}
\authoremail{gfuller@ucsd.edu, shi@physics.ucsd.edu}

\begin{abstract}
We propose that the gravitational collapse of
supermassive objects (${\rm M}\ga 10^4\,{\rm M_\odot}$),
either as relativistic star clusters or as
single supermassive stars (which may result from
stellar mergers in dense star clusters),
could be a cosmological source of $\gamma$-ray bursts.
These events could provide the seeds of the
supermassive black holes observed at the center
of many galaxies. Collapsing supermassive objects will
release a fraction of their huge gravitational
binding energy as thermal neutrino pairs.
We show that the accompanying
neutrino/antineutrino annihilation-induced heating
could drive electron/positron \lq\lq
fireball\rq\rq\ formation, relativistic expansion,
and associated  $\gamma$-ray emission.
The major advantage of this model is its energetics:
supermassive object collapses are far more
energetic than solar mass-scale compact object mergers;
therefore, the conversion of gravitational energy
to fireball kinetic energy in the supermassive object
scenario need not be highly efficient, nor is it
necessary to invoke directional beaming. The major weakness of
this model is difficulty in avoiding a baryon loading problem
for one dimensional collapse scenarios.
\end{abstract}


\section{Introduction}

In this letter we propose that the collapse of supermassive
objects and the associated neutrino/antineutrino annihilation
could give rise to high redshift (cosmological) $\gamma$-ray
bursts (GRBs). This model could alleviate vexing problems
associated with the energetics
of conventional stellar remnant-based scenarios.
We define a supermassive object to be a star or star cluster
that suffers the general relativistic
Feynman-Chandrasekhar instability during its evolution.
This corresponds to objects with initial masses
$M \ga 10^4\,M_\odot$, i.e., those which may
leave black hole remnants with masses $M \ga 10^3\,M_\odot$.

Detections of absorption and emission
features at a redshift $z=0.835$ in the spectral observation
of the afterglow of $\gamma$-ray burst
GRB970508 (Metzger et al. 1997a,b) and at redshift
$z=3.42$ in the host galaxy of GRB971214 (S. Kulkarni
{\it et al.} 1998) have established that
at least some of the GRB sources lie at
cosmological distances. Observations show that the
total energy in gamma rays associated with
a GRB at cosmological distances is
$\sim  {10}^{52}\,{\rm erg}$ to $\sim {10}^{53}\,{\rm erg}$
when a $4\pi$ solid angle coverage is assumed
(Fenimore et al. 1993; Wijers et al. 1997;
Kulkarni {\it et al.} 1998).
Catastrophic collapse events,
such as neutron-star/neutron-star mergers (Paczy\'nski 1986;
Goodman 1986; Eichler et al. 1989), neutron-star/black-hole
mergers (Mochkovitch et al. 1993), failed supernovae
(Wooseley 1993), \lq\lq hypernovae\rq\rq\ (Paczy\'nski 1997),
collapse of Chandrasekhar-mass white
dwarfs (Usov 1992), have been touted as natural candidates for
cosmological GRB sources.  Fireballs created in
these collapse events could accelerate material to the
ultra-relativistic regime, with Lorentz factors $\Gamma =
E_e/m_ec^2\ga 10^2$ (Paczy\'nski 1986, Goodman 1986, Rees \&
M\'esz\'aros 1992, M\'esz\'aros \&  Rees 1992).
The kinetic energy in these fireballs could then be converted
to $\gamma$-rays possibly via the cyclotron radiation and/or
the inverse Compton processes associated with
ultrarelativistic electrons.
In these models, the energy loss of the shock(s) propelled
by the fireball would produce the afterglow associated with
a GRB event (Waxman 1997).

There are, however, problems for these stellar
remnant-based models if the GRBs originate from high
redshift events. The total gravitational
binding energy released when a $\sim 1\,{\rm M}_\odot$
configuration collapses to a black hole (or into a
pre-existing larger black hole) is only
$\sim10^{54}$ erg.  Calculations have shown
that it is very difficult to power a GRB of
energy $\sim 10^{52}$ erg  (Wijers et al. 1997), or an
afterglow with a similar energy (Waxman 1997; Dar 1997)
with such a collapse scenario, unless the $\gamma$-ray
emission and the blast wave causing the
afterglow are highly collimated (improbably
highly collimated in the case of
very high redshift events).

This energetics problem can be avoided in the
supermassive object collapse
model suggested here. Collapse of such large mass scale 
objects could result in prodigious gravitational binding
energy release. Some of this gravitational energy is radiated
as thermal neutrino/antineutrino pairs
(Fuller, Woosley, \& Weaver 1986,
hereafter FWW; Fuller \& Shi 1997)
whose annihilations into electron/positron pairs could
create a fireball above the core that generates $\gamma$-rays.
There is no direct evidence for supermassive stars
ever having been extant in the universe.
However, it has been argued that their formation could be an
inevitable result of the collapse of
$\sim 10^5{\rm M}_\odot$ to $10^6{\rm M}_\odot$ primordial
clouds (the baryon Jean's mass at early epochs, see
Peebles \& Dicke 1968, and Tegmark et al. 1997)
at high redshifts in which cooling was not as
efficient as in clouds contaminated with metals,
or more likely, as a result of stellar mergers associated
with $\ga 10^7$--$10^8M_\odot$ relativistic star cluster
collapse (Hoyle \& Fowler 1963; Begelman \& Rees 1978;
Bond, Arnett, \& Carr 1984; FWW; McLaughlin \& Fuller 1996).
The flow chart for supermassive black hole production
suggested by Begelman \& Rees (1978) includes several
pathways whereby supermassive stars are formed in the
central region of the collapsing cluster.
Further, supermassive black holes apparently
are ubiquitous in the universe.
They are invoked as the central engines of Active
Galactic Nuclei (AGNs) and quasars, and are inferred to be in
the centers of nearby galaxies (van der Marel et al. 1997).

We note that Prilutski and Usov (1975)
have previously tied GRBs to
magneto-energy transfer during collapses of
supermassive rotators ($\sim 10^6{\rm M}_\odot$)
postulated to power AGNs and quasars.
Here we propose a different energy transfer mechanism
(neutrinos) based on objects not necessarily tied to
AGNs or quasars, but which could possibly be
related to the birth of the supermassive
black holes that power them.

\section{Fireballs from Supermassive Object Collapse}

Supermassive stars will suffer the
General Relativistic (Feynman-Chandrasekhar)
instability, either at or before the onset of hydrogen burning
(c.f., FWW) in the case of quasi-statically contracting objects,
or immediately upon formation
as in the case where stellar mergers produce them.
As such a star collapses, the entropy per baryon is
slightly increased by nuclear burning,
but then is reduced by neutrino pair emission.  Though
initially the whole star can collapse homologously, as
the entropy is reduced only an inner \lq\lq homologous core\rq\rq\
can continue to collapse homologously (FWW).
It is this homologous core that will plunge through an event
horizon as a unit to make a black hole.  The mass of the
homologous core, $M_5^{\rm HC}\equiv M^{\rm HC}/10^5M_\odot$,
can be much smaller (possibly by an order of magnitude or more)
than the mass of the initial hydrostatic supermassive star,
$M_5^{\rm init}\equiv M^{\rm init}/10^5M_\odot$.

The collapse to a black hole of a supermassive star with a
homologous core mass $M^{\rm HC}$ will have a characteristic
(prompt) Newtonian gravitational binding energy release of
$\sim E_{\rm s}\approx 10^{59}M^{\rm HC}_5\,{\rm erg}$.
During the collapse, neutrino emission will ensue from
$e^\pm$-annihilation in the core. The emissivity of this process
scales as the core temperature to the {\it ninth}
power (Dicus 1972). As a result, most of the
gravitational binding energy removed by neutrinos will be
emitted very near the point where the core becomes a black
hole, and on a timescale characterized by the free fall time
(or light crossing time) of the homologous core near the
black hole formation point.  We employ a characteristic free fall
collapse time scale of $t_{\rm s}\approx M^{\rm HC}_5\,{\rm sec}$,
and a characteristic radius (the Schwarzschild radius)
of $r_{\rm s}\approx 3\times 10^{10}M^{\rm HC}_5\,{\rm cm}$.
For a core mass $\ga 10^4\,{\rm M}_\odot$ the neutrinos will
not be trapped in the core and will freely stream out.
For a smaller core mass, the neutrino diffusion timescale
will be long compared to the free fall
timescale and so neutrinos will be trapped in the core.
Neutrino emission in this latter case will be from a \lq\lq
neutrino sphere\rq\rq\ at the edge of the homologous core.

In general it is difficult to estimate the range of initial
stellar masses that will give rise to a given range of
homologous core masses, though there is a
clear hierarchy at each evolutionary stage. We therefore guess
that the initial star cluster masses will be in the range
${10}^{5}\,{\rm M}_{\odot}$ to ${10}^{9}\,{\rm M}_{\odot}$,
while the subsequently produced supermassive stars
will have masses $M_5^{\rm init} \approx 0.1$ to
$\sim 1000$, while the corresponding homologous core
masses will lie in the range $M^{\rm HC}_5
\approx {10}^{-2}$ to $\sim 10$. Figure 1 shows a flow chart
for the collapse of supermassive objects.

The neutrino luminosity can be crudely estimated from the
product of the neutrino energy emissivity (Schinder et al.
1987; Itoh et al. 1989) near the black hole formation point
and the volume inside the Schwarzschild radius, i.e.,
$4\times 10^{15}\,(T_9^{\rm Schw})^9\,(4\pi r_{\rm s}^3/3)$
erg/sec.  Here $T_9^{\rm Schw}$ is the characteristic
{\it average} core temperature near the black hole formation
point in units of $10^9$ K.
For a spherical non-rotating supermassive star we can show that
\begin{equation}
T_9^{\rm Schw} \approx 12 {\alpha}_{\rm Schw}^{1/3}
{\left({{11/2}\over{g_{\rm s}}}\right)}^{1/3} {\left( {{M_5^{\rm
init}}\over{M_5^{\rm HC}}}\right)}^{1/6} {\left( M_5^{\rm HC}
\right)}^{-1/2},
\label{temp}
\end{equation}
where ${\alpha}_{\rm Schw}$ is the ratio of the final entropy
per baryon to the value of this quantity in the initial
pre-collapse hydrostatic configuration,
and $g_{\rm s} \approx g_b + 7/8 g_f \approx 11/2$ is the
statistical weight of relativistic particles in the core.
Since for spherical non-rotating supermassive stars
$M_5^{\rm init}/M_5^{\rm HC}\approx \sqrt{5.5/2}{\alpha}_{\rm  
Schw}^{-2}$ (FWW), we can conclude that
$T_9^{\rm Schw}\approx 13(M^{\rm HC}_5)^{-1/2}$.
The characteristic neutrino luminosity is then
\begin{equation}
L_{\nu\bar\nu}\sim 4\times 10^{15}\,(T_9^{\rm Schw})^9\,(4\pi\,
r_{\rm s}^3/3)\,{\rm erg/sec}
\approx 5\times 10^{57} (M^{\rm HC}_5)^{-3/2}\,{\rm erg/sec}.
\label{luminosity}
\end{equation}
Since 70\% of the neutrino emission is in the
$\nu_e\bar\nu_e$ channel, the characteristic 
luminosity of $\nu_e$ or $\bar\nu_e$ is
$L_{\nu_e}=L_{\bar\nu_e}\approx 0.35L_{\nu\bar\nu}$.
(This estimate of $L_{\nu\bar\nu}$
is a factor of $\sim 10$ above an appropriately scaled version of
the Woosley, Wilson and Mayle (1986) result for a
$M_5^{\rm init} = 5$ configuration; part of the difference
is attributable to the employment of different neutrino
emissivities, and the remainder may result from
different core temperatures.)

The copious $\nu\bar\nu$ emission during the collapse can
create a fireball above the homologous core by
$\nu\bar\nu\rightarrow e^+e^-$.
Clearly, the neutrino luminosities will suffer gravitational
redshift which will degrade the total energy deposition
above the star, though this will be compensated by increased
$\nu\bar\nu$-annihilation from gravitational bending
of null trajectories (Cardall \& Fuller 1997).  A detailed
calculation of these two effects is beyond the scope of
this paper, but we do not expect the combination of them
to change our order-of-magnitude estimates significantly.
The energy deposition rate per unit volume
from the $\nu\bar\nu$ annihilation at a radius $r$
above a spherical shell of thermal neutrino emission
with a radius $R_\nu$, is then
(Goodman, Dar, \& Nussinov 1987;
Cooperstein, van den Horn, \& Baron 1987)
\begin{equation}
\dot{Q}_{\nu\bar\nu}(r)={KG_F^2\Phi(x)\hbar^2\,c\over  
12\pi^2\,R_\nu^4}
                         L_\nu\,L_{\bar\nu}
			\Bigl[{\langle E_\nu^2\rangle\over
			       \langle E_\nu\rangle}+
			      {\langle E_{\bar\nu}^2\rangle\over
			       \langle E_{\bar\nu}\rangle}\Bigr].
\end{equation}
Here $G_F$ is the Fermi constant, $L$ is the luminosity of the
neutrinos/antineutrinos, and the brackets denote averages of
neutrino energy or squared-energy over the appropriate neutrino
or antineutrino energy spectra (see Shi \& Fuller 1998).
The phase space and spin factors are
$K\approx0.124\,\,(0.027)\quad {\rm for\ }\nu=\nu_e\,
(\nu_\mu{\rm ,}\nu_\tau)$,
and the radial dependence of the energy deposition rate is
$\Phi(x)=(1-x)^4\,(x^2+4x+5)$, with $x=[1-(R_\nu/r)^2]^{1/2}$.

The characteristic neutrino luminosity $L_{\nu\bar\nu}$
in eq.~(\ref{luminosity}) could be
an underestimate of the true neutrino luminosity.
A detailed numerical calculation (without
considering the uncertain gravitational redshift, however)
shows that the true average neutrino luminosity
can be much higher if there is rapid rotation and/or
magnetic fields holding up the collapse (Shi \& Fuller 1998).
The neutrino energy loss rate scales steeply
as $T_9^9$, and the temperature distribution in the
homologously collapsing core (an index $n=3$ polytrope)
follows the Lane-Emden function and so peaks at the center.
Compensating this feature will be the $R_\nu^4$ dependence of
the above $\nu\bar\nu$ energy deposition rate $\dot{Q}_{\nu\bar\nu}$.
Therefore, we will approximate the
entire neutrino emissivity of the core as arising from the
edge of the core ($R_\nu\sim r_{\rm s}$), and then take
$L_{\nu\bar\nu}$ as the characteristic neutrino luminosity
from eq.~{(\ref{luminosity}). (Note that this equation
is appropriate in the case where $M_5^{\rm HC} \la
%{\\lower-1.2pt\vbox{\hbox{\rlap{$<$}\lower5pt\vbox{\hbox{$\sim$}}}}\ }  
0.1$ and neutrinos diffuse from the core. In this case,
the central temperature is irrelevant, though we may get
luminosities comparable to the free streaming
case because the core will have lower mass and, hence,
a generally higher temperature scale.)

The expected near-thermal spectrum of the neutrino emission implies
$\langle E_\nu^2\rangle/\langle E_\nu\rangle=
\langle E_{\bar\nu}^2\rangle/\langle E_{\bar\nu}\rangle
\approx 6\,(M^{\rm HC}_5)^{-1/2}\,{\rm MeV}$ (Shi \& Fuller 1998).
Therefore, the neutrino energy deposition rate per unit volume
will be roughly
\begin{equation}
\dot{Q}_{\nu\bar\nu}(r)\sim 4\times 10^{22}\,(M^{\rm HC}_5)^{-7.5}
(r_{\rm s}/r)^8\,{\rm erg}\,{\rm cm}^{-3}{\rm s}^{-1}.
\end{equation}
The total energy deposited into the fireball above a radius $r$ is
\begin{equation}
E_{\rm f.b.}(r)=t_{\rm s}\,
{\int_r^\infty} 4\pi r^2\dot{Q}_{\nu\bar\nu}(r){\rm d}r
\sim 2.5\times 10^{54}\,(M^{\rm HC}_5)^{-3.5}(r_{\rm s}/r)^5\,
{\rm erg},
\label{fireball}
\end{equation}
which is tremendous.  The fireball will undoubtedly lose some of
this energy to thermal neutrino emission. But, once the $e^\pm$ 
pair density is high enough for this, neutrino/electron
scattering should deposit even more energy. If
$M_5^{\rm HC}=0.5$, the energy deposited in the fireball
will be $\sim10^{53}$ erg at a radius $r\sim 3r_{\rm s}
\sim 10^{11}\,{\rm cm}$.  This is the total
observed energy in a GRB assuming a $4\pi$ solid angle  
and a redshift $z\sim 3$.

A successful model of GRBs must avoid
excessive baryon loading so that a Lorentz factor of
$\Gamma \ga 10^2$ can be achieved for the baryons accelerated
by the fireball. This suggests that the region at several
Schwazschild radii from the supermassive star core should have
extremely low baryon density.  This may be satisfied if
the {\it whole} star collapses homologously into a black hole,
and/or substantial rotation causes the star to collapse in a
flattened geometry with very little material in the
polar directions (an extreme case of this geometry was
discussed in Bardeen \& Wagoner 1969).  The homologous
collapse of the entire star could be engineered only if
the star has substantial centrifugal support
from rotation and/or if there is significant magnetic
pressure (but not so much that an explosion results).
Therefore, rotation could be a crucial factor in this picture.
Rotation will also result in a longer collapse timescale,
and mildly beamed $\gamma$-ray emission.  A high angular
momentum collapse may therefore be challenged in generating GRBs
with durations $\la 1$ second.

Another means to avoid excessive baryon loading may be
possible in the collapse of a dense star cluster.
In this case the whole star cluster can
collapse on the General Relativistic instability
(Shapiro \& Teukolsky 1985) and 
collisions of $M_*\sim M_\odot$ stars could provide the
neutrino \lq\lq engine\rq\rq\ that powers fireballs.
During the collapse, the central stars will have
relativistic speeds and the typical entropy per baryon
produced in zero impact parameter collisions of these 
will be $S\sim 10^4\Gamma^{1/2}(g_s/5.5)^{1/4}
(M_\odot/M_*)^{1/4}(V_*/V_\odot)^{1/4}$ with $T_9\sim 1$,
conditions commensurate with those required for hydrostatic
supermassive stars 
($S\approx 10^4(M^{\rm init}/10^8M_\odot)^{1/2}$).
(Here $\Gamma \sim 1$ is an appropriate Lorentz factor, and
$V_*/V_\odot$ is the ratio of the stellar collision
interaction volume to the solar volume.) In fact, most
collisions will not be \lq\lq head-ons,\rq\rq\
but rather will involve the tenuous outer layers of the stars.
The lower densities involved will translate into larger
entropies (effectively, $(V_*/V_\odot)^{1/4}$ could be 
considerably larger), possibly large enough
($S\sim {10}^7$) to provide a pair fireball directly.
In the collapse, space between moving stars may provide
baryon-free \lq\lq lanes\rq\rq, and the stellar collisions
themselves may cause the neutrino emission
to be \lq\lq spiky\rq\rq\ (the overall emission profile,
however, should nevertheless follow the free fall collapse
profile indicated above for supermassive stars).
Both processes are stochastic, possibly contributing to the
\lq\lq spiky\rq\rq\ time structure of the GRBs.
This direct collapse of relativistic star clusters and
the collapse of supermassive stars may well represent
two extremes on a continuum of supermassive object collapse.

\section{Event Rate and Peak Flux Distribution}

The rate of supermassive object collapses should be able
to match the observed rate of GRB events
(several per day) if a substantial fraction of the burst
events are to come from this source.
Assuming that supermassive objects all form and collapse
at a redshift $z$, the rate of these collapses as observed
at the present epoch is
\begin{equation}
4\pi r^2a_z^3{{\rm d}r\over {\rm d}t_0}{\rho_{\rm b}F(1+z)^3
\over M^{\rm init}},
\end{equation}
where $r$ is the Friedman-Robertson-Walker comoving coordinate
distance of the objects, $a_z$ is the scale factor of the
universe at the epoch corresponding to a
redshift $z$ (with $a_0=1$), $t_0$ is the age of the universe,
$\rho_{\rm b}\approx 2\times 10^{-29}\,\Omega_{\rm b}h^2\,{\rm
g}\,{\rm cm}^{-3}
\approx 5\times 10^{-31}{\rm g}\,{\rm cm}^{-3}$ 
(Tytler \& Burles 1997)
is the baryon density of the universe today, $h$ is the
Hubble parameter in 100 km$\,{\rm s}^{-1}\,{\rm Mpc}^{-1}$,
and $F$ is the fraction of baryons that were incorporated 
in supermassive objects.  For $z\sim 3$ we will have
$r\sim 3000h^{-1}$ Mpc. The collapse rate is therefore
\begin{equation}
0.15F\,(M_5^{\rm init})^{-1}\,{\rm sec}^{-1}
\sim 10^4F\,(M_5^{\rm init})^{-1}\,{\rm day}^{-1}.
\end{equation}
With $F\sim 0.1\%$, i.e., with 0.1$\%$ of all baryons having
been incorporated into supermassive objects of
$M_5^{\rm init}\sim 10$, we should observe
(assuming a 100\% detection efficiency)
one collapse per day if they emitted $\gamma$-rays into
a $4\pi$ solid angle. This would constitute a substantial
fraction of the observed rate of GRB events.
The baryon fraction $F=0.1\%$ in $\sim 10^6 {\rm M}_\odot$
black holes implies a (cumulative) density of 7$h^2$ such
supermassive black holes formed in 1 Mpc$^3$.  This
GRB rate is about two orders of magnitude lower than
24 Gpc$^{-3}\,{\rm yr}^{-1}$, the rate required if GRBs originate
from source populations that do not evolve over time
(Fenimore and Bloom 1995). This shortfall in rate results
because we have assumed that all GRBs are high redshift 
collapse events and are therefore seen from a larger volume.
In addition, the rate of supermassive object collapses
required in our GRB model does not depend on the mass 
scale of the collapsing objects, although the fraction
$F$ scales linearly with $M_5^{\rm init}$.
Observations show that almost all galaxies that have been
examined appropriately seem to have supermassive black holes
in their centers (van den Marel et al. 1997). 
It is therefore intriguing to estimate the rate of
supermassive object collapses required
by our GRB model on a per galaxy basis.
If such supermassive object collapses
occurred only in normal $\sim L_*$ galaxies, the
rate needed is about 350$h^{-1}$ per $L_*$ galaxy. However,
this number of events per galaxy is much lower,
perhaps $\la 10h^{-1}$ per galaxy
(based on, for example, the galaxy number densities of 
Zucca et al. 1997), if dwarf galaxies harbor supermassive
objects as well. Therefore, it may be conceivable that these
supermassive object collapse events are tied to the supermassive
black holes at the centers of galaxies, if such supermassive
black holes occur in every galaxy-scale object. 
Such an association of supermassive objects and galaxy-scale
objects may also be born out by considering Lyman limit systems
and damped Lyman-$\alpha$ systems, which are associated with
galactic halos and disks at high redshifts.  Using a column 
density $N_{\rm HI}$ distribution per unit column density per
unit absorption distance of $10^{13.9}N_{\rm HI}^{-1.74}$
(Storrie-Lombardi, Irwin \& McMahon 1996),
we find that the rate of supermassive object collapse matches
that of GRBs if every Lyman-$\alpha$ system with $N_{\rm HI}
\ga 10^{18}\,{\rm cm}^{-2}$ harbors a supermassive object.

If all GRBs are from $z\ga 1$ then the $\gamma$-ray burst
peak flux distribution ($\log N$-$\log P$) will be very
different from models with a homogeneously distributed 
population of GRBs.
The observed $\log N$-$\log P$ distribution is a power law
with index $=-1.5$ which has a break at the faint end
(Fenimore et al. 1993). This would be consistent with
homogeneously distributed cosmological sources with a
cut-off at high redshifts, unless the peak flux of GRBs,
$P$, cannot be regarded as a standard candle.
But since the $\log N$-$\log P$ distribution is a
convolution of the peak flux and spatial distribution,
there is no guarantee that the observed
power law requires a homogeneous distribution
of sources. For our model, in which supermassive object
collapses most likely occur at cosmological distances with
$z\ga 1$, we can always invoke variances in the peak
flux of GRBs, and/or an evolution of supermassive object
co-moving number densities, or invoke another population
of GRBs, to fit the observed $\gamma$-ray burst peak flux
distribution. It is worth noting that even in existing
stellar remnant-based models, the sources tend to be
more abundant at $z\ga 1$, because the star formation rate was
higher then (Totani 1997).

\section{Conclusion}

The formation of the supermassive black holes
inferred in AGNs, quasars and many galaxies may
well involve the collapse of relativistic star clusters
which form intermediate phase supermassive stars.
We point out here that collapses of these supermassive objects
will be accompanied by prodigious thermal neutrino emission which
could transport a fraction of the gravitational binding
energy of these objects to a region(s) where the baryon loading
is low, thus creating \lq\lq clean\rq\rq\
fireballs that generate $\gamma$-ray bursts.
The major advantage of this model is a 
huge energy release, and just such an energy scale is
required by recent observations of high redshift bursts.
We have shown that the collapse timescale and
expected collapse event rates are consistent with
$\gamma$-ray burst parameters.  The principal weakness
of our model is the baryon loading problem.  We have
outlined possible ways to circumvent this problem by
appealing to high angular momentum and flattened collapses,
and by appealing to the stochastic nature of stellar
collision-induced supermassive star/black hole build-up in the
collapse of relativistic star clusters.

We thank David Band and Edward Fenimore
for valuable suggestions.  This work is supported
by NASA grant NAG5-3062 and NSF grant PHY95-03384 at UCSD.

\newpage
\begin{center}
{\bf Figure Caption}
\end{center}
\smallskip

\noindent Figure 1. A flow chart for the collapse
of supermassive objects.
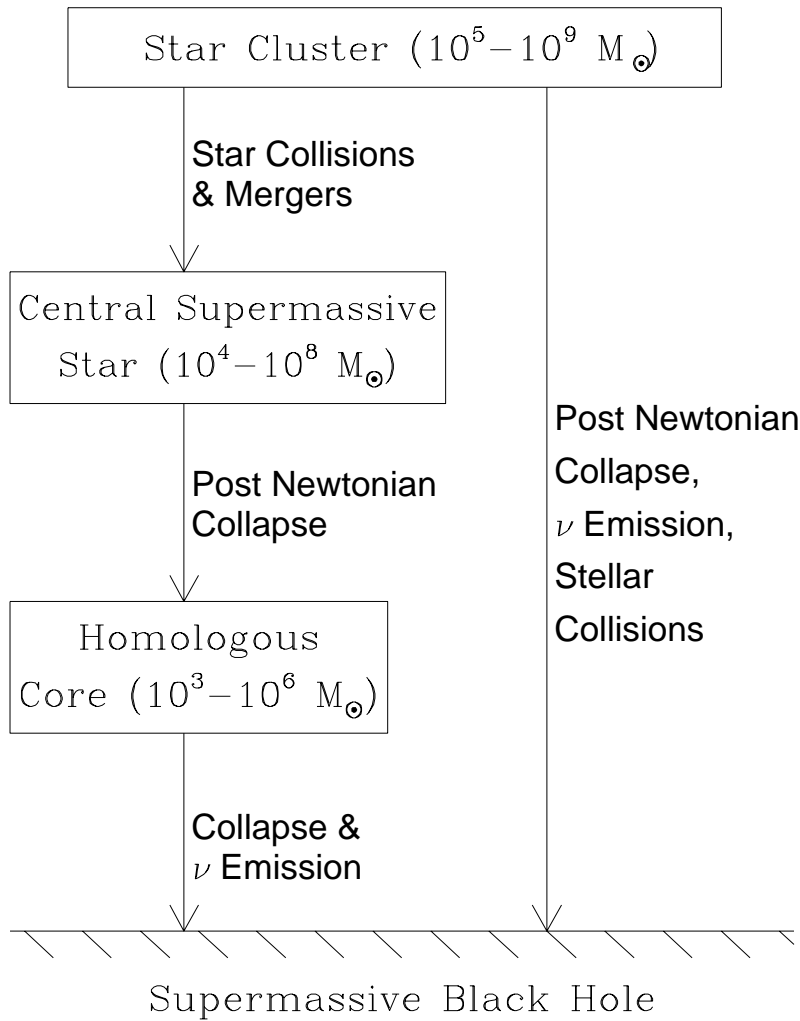
\end{document}